\documentclass[oribibl]{llncs}
\usepackage{multicol}
\usepackage{amsmath}
\usepackage{amssymb}
\usepackage{graphicx}
\usepackage{color}
\usepackage{hhline}
\usepackage{pdfsync}
\newcommand{\ignore}[1]{}

\newcommand{\boxtheorem}{\hfill $\Box$}
\newcommand{\nit}[1]{{\it #1}}

\usepackage{graphicx}

\usepackage{algpseudocode}
\usepackage{algorithm}

\newcounter{lemmaA-counter}
\newcounter{propositionA-counter}
{\vskip \abovedisplayskip \refstepcounter{lemmaA-counter}%
\noindent {\bf Lemma A.\arabic{lemmaA-counter}.}}%

{\vskip \abovedisplayskip \refstepcounter{propositionA-counter}%
\noindent {\bf Proposition A.\arabic{propositionA-counter}.}}%

\def\verbatim@font{\rmfamily\small}
\makeatother

\newcommand{\mc}[1]{\mathcal{ #1}}

\newcommand{\ignoreT}[1]{}

\newcommand{\nc}{{\em nc}}

\newcommand{\dpm}{{Datalog}$^\pm$}

\newcommand{\schema}{\mc{R}}

\newcommand{\dplus}{{Datalog}$^+$}

\newcommand{\omd}{OMD}

\newcommand{\hm}{HM}

\newcommand{\fd}{FD}

\newcommand{\ideps}{IDs}

\newcommand{\egds}{{\em egds}}
\newcommand{\egd}{{\em egd}}
\newcommand{\tgds}{{\em tgds}}
\newcommand{\tgd}{{\em tgd}}
\newcommand{\ncs}{{\em ncs}}

\title{\vspace*{-2cm} {\bf The Ontological Multidimensional Data Model}\\\vspace{-3mm}{\small (extended abstract)}\vspace{-7mm}}
\author{{\bf Leopoldo Bertossi}\thanks{ Carleton Univ., \
School of Computer Science, Canada. \
bertossi@scs.carleton.ca} \ and \ {\bf Mostafa Milani}\thanks{ McMaster Univ., Dept. Computing and Software, Canada. \
mmilani@mcmaster.ca }}

\institute{}

\begin{document}
\pagestyle{plain}
\maketitle
\thispagestyle{empty}

\begin{abstract}
We briefly present {\em OMD}, a model of multidimensional data that uses  ontologies written in \dpm, an extension of the classical declarative language Datalog  for relational databases.

\end{abstract}

\vspace{1mm}
We present the {\em Ontological Multidimensional Data Model} (\omd) as an ontological, \dpm-based \cite{AC09} extension of the Hurtado-Mendelzon (\hm) \ model for multidimensional data \cite{hurtado-pods}.

For limitations of space, we will use  a running example  to illustrate the main elements of an \omd \ model.

\vspace{-3mm}
\begin{figure}[ht]
\begin{center}
\vspace{-4mm}
 \includegraphics[width=9.5cm]{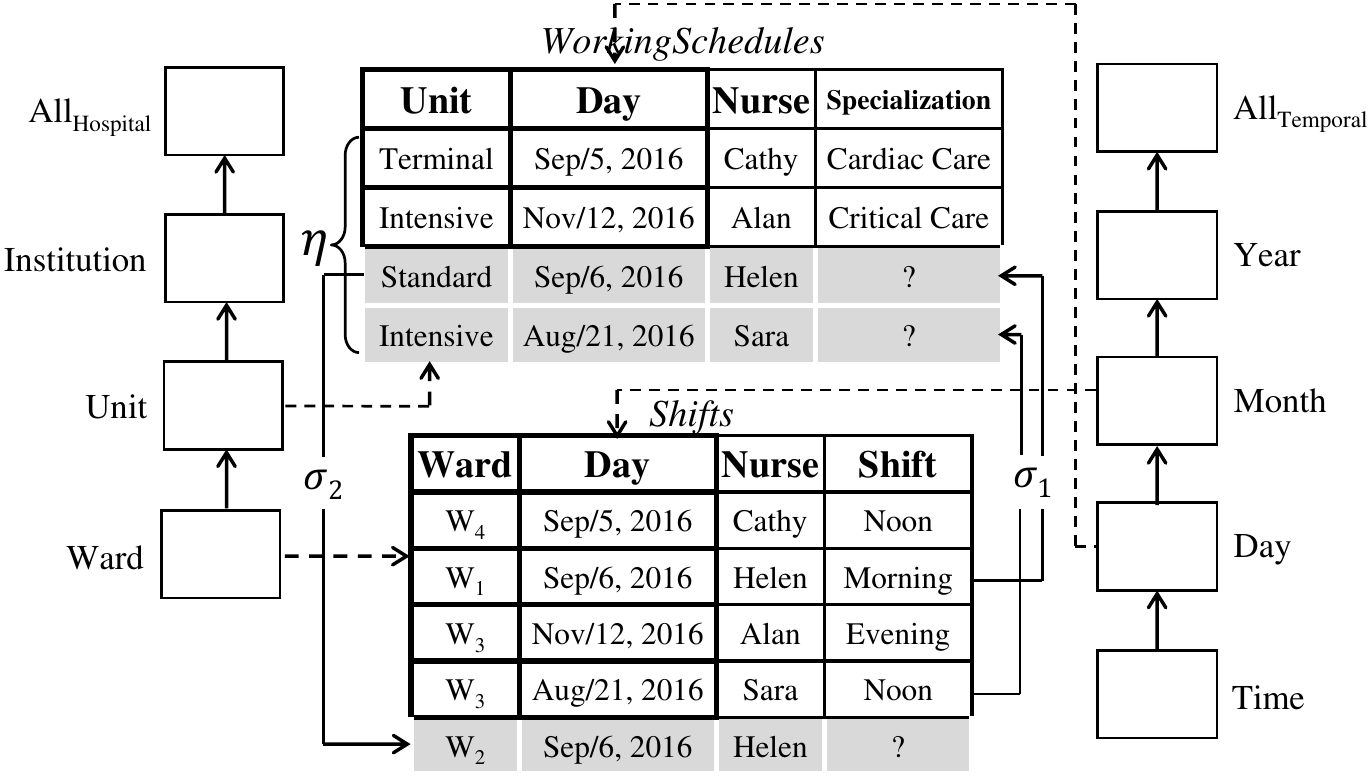} \vspace{-3mm}
 \caption{An \omd \ model with categorical relations, dimensional rules, and constraints}\label{fig:omdm} \vspace{-10mm}
\end{center}
\end{figure}

An OMD model has a  {\em database schema}  $\schema^\mc{M}=\mc{H} \cup \mc{R}^c$, where $\mc{H}$ is a relational schema with multiple dimensions, with sets $\mc{K}$ of unary category predicates,  and sets $\mc{L}$ of binary, child-parent predicates; and $\mc{R}^c$ is a set of {\em categorical predicates}.


\vspace{2mm}
\hspace*{-5mm}\begin{minipage}[t]{0.55\textwidth}
\noindent {\em Example:}  Figure \ref{fig:omdm} shows $\sf{Hospital}$ and $\sf{Temporal}$  dimensions. The former's instance is here on the RHS.  $\mc{K}$ contains  predicates $\nit{Ward}(\cdot)$, $\nit{Unit}(\cdot)$, $\nit{Institution}(\cdot)$, etc. Instance $D^\mc{H}$ gives them extensions, e.g.  $\nit{Ward}$ $=$ $\{{\sf W}_1,{\sf W}_2,{\sf W}_3,{\sf W}_4\}$.  $\mc{L}$ contains, e.g. $\nit{WardUnit}(\cdot,\cdot)$, with extension: \   $\nit{WardUnit}$ $=$ $\{({\sf W}_1,$ $ {\sf standard}),$ $({\sf W}_2,$ ${\sf standard}),$ $({\sf W}_3,$ ${\sf intensive}),$
  $({\sf W}_4,$ ${\sf terminal})\}$. \
In the middle of Figure \ref{fig:omdm},  {\em categorical relations} are associated to dimension categories. \boxtheorem
\end{minipage}

\vspace{-5.8cm}

\begin{center}
\begin{figure}
\hspace*{7.4cm}\includegraphics[width=4.5cm]{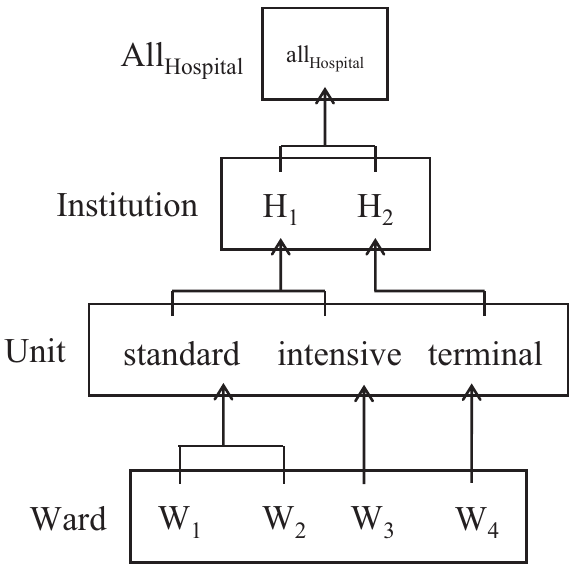}
\end{figure}
  \end{center}

\vspace{-1cm}
Attributes of categorical predicates are either {\em categorical}, whose values are members of dimension categories, or {\em non-categorical}, taking values from arbitrary domains. Categorical predicate are represented in the form  $R(C_1,\ldots,C_m;N_1, \ldots,$ $N_n)$, with categorical attributes before ``;'' and non-categorical  after.

The extensional data, i.e the instance for the schema $\schema^\mc{M}$, is $I^\mc{M}=D^\mc{H}\cup I^c$, where $D^\mc{H}$ is a complete instance for dimensional subschema $\mc{H}$ containing the
category and child-parent predicates; and  sub-instance $I^c$ contains possibly partial, incomplete extensions for the categorical predicates, i.e. those in $\mc{R}^c$.

 Schema $\schema^\mc{M}$  comes with basic, application-independent semantic constraints, listed below.

\vspace{1mm}
\noindent {\bf 1.} \  Dimensional child-parent predicates must take their values from categories. Accordingly, if child-parent predicate $P \in \mc{L}$ is associated to category predicates $K,K' \in \mc{K}$, in this order, we
introduce inclusion dependencies (\ideps) as \dpm \ {\em negative constraints} (\ncs): \
$P(x,x'), \ \lnot K(x) ~\rightarrow~ \bot, \  \mbox{ and } \
P(x,x'), \ \lnot K'(x') ~\rightarrow~ \bot$.
\ (The $\bot$ symbol denotes an always false propositional atom.) \
We do not represent them as \dpm's {\em tuple-generating dependencies} (\tgds) \ $P(x,x') \rightarrow K(x)$, etc., because we reserve  \tgds \ for possibly incomplete predicates (in their RHSs).

\vspace{1mm}
\noindent {\bf 2.} \ Key constraints on dimensional child-parent predicates $P \in \mc{K}$, as {\em equality-generating dependencies} (\egds): \ $P(x,x_1),P(x,x_2) ~\rightarrow~ x_1=x_2$.

\vspace{1mm}
\noindent {\bf 3.} \
The connections between categorical attributes and the category predicates are specified by means of  \ncs. For categorical predicate $R$, the \nc \
$R(\bar{x};\bar{y}), \ \lnot K(x)$ $\rightarrow~ \bot$,
where $x \in \bar{x}$ takes values in category $K$.

\vspace{1mm}
 \noindent {\em Example:}  The categorical attributes {\it Unit} and {\it Day} of categorical predicate \linebreak $\nit{WorkingSchedules}(\!\nit{Unit},\!\nit{Day};\!\nit{Nurse}\!,\nit{Speciality})$ in $\mc{R}^c$ \ignore{and $\nit{Shifts}(\!\nit{Ward},\!\nit{Day};\!\nit{Nurse},\!\nit{Shift})$ in Figure~\ref{fig:omdm} are categorical relations. In \nit{WorkingSchedules}, {\it Unit} and {\it Day} are categorical attributes,} are connected to the {\sf Hospital} and {\sf Temporal} dimensions, resp., as captured by the  \ideps \ $\nit{WorkingSchedules}[1]\subseteq \nit{Unit}[1]$, and $\nit{WorkingSchedules}[2]\subseteq \nit{Day}[1]$. The former is written in \dplus as \
${\it WorkingSchedules(u,d;n,t)},\lnot {\it Unit}(u)  ~\rightarrow~ \bot$.
For the {\sf Hospital} dimension,  one of the  \ideps \ for  predicate $\nit{WardUnit}$ is $\nit{WardUnit}[2] \subseteq \nit{Unit}[1]$, which  is expressed by the \nc:
\ ${\it WardUnit(w,u)},\lnot {\it Unit}(u)$  $\rightarrow \bot$. \
 The key constraint of \nit{WardUnit} is captured by the \egd: \ ${\it WardUnit(w,u)},$ ${\it WardUnit(w,u')}  ~\rightarrow$ $u=u'$. \boxtheorem

\vspace{1mm}
The \omd \ model allows us to build  {\em multidimensional ontologies}, $\mc{O}^\mc{M}$. In addition to an instance  $I^\mc{M}$ for a schema $\mc{R}^\mc{M}$, they include the set $\Omega^\mc{M}$ of {\em basic constraints} as in {\bf 1.}-{\bf 3.} above, a  set $\Sigma^\mc{M}$ of {\em dimensional rules} (those in {\bf 4.} below), and a set $\kappa^\mc{M}$ of {\em dimensional constraints} (in {\bf 5.} below); all of them application-dependent and expressed in the \dplus \ language associated to schema $\schema^\mc{M}$.

\vspace{1mm}
\noindent {\bf 4.} \ {\em Dimensional rules} as \dplus \ \tgds: \ $R_1(\bar{x}_1;\bar{y}_1),...,R_n(\bar{x}_n;\bar{y}_n),P_1(x_1,x'_1),...,$ $P_m(x_m,x'_m) \ \rightarrow \ \exists \bar{y}' \ R_k(\bar{x}_k;\bar{y})$. \
Here, the $R_i(\bar{x}_i;\bar{y}_i)$) are categorical predicates, the $P_i$ are child-parent predicates, $\bar{y}' \subseteq \bar{y}$, \ $\bar{x}_k \subseteq \bar{x}_1 \cup ... \cup \bar{x}_n \cup \{x_1,...,x_m, x'_1,...,x'_m\}$, \ $\bar{y} \! \smallsetminus \! \bar{y}' \subseteq \bar{y}_1 \cup ... \cup \bar{y}_n$; repeated variables in bodies (join variables) appear only categorical positions in categorical relations and  in child-parent predicates. Existential variables appear only in non-categorical attributes.

\vspace{1mm}
\noindent {\bf 5.} \
{\em Dimensional constraints}, as \egds \ or \ncs: \
$R_1(\bar{x}_1;\bar{y}_1),...,R_n(\bar{x}_n;\bar{y}_n),P_1(x_1,x'_1),$ $...,P_m(x_m,x'_m) \rightarrow z=z'$, \ and \ $R_1(\bar{x}_1;\bar{y}_1),...,R_n(\bar{x}_n;\bar{y}_n), P_1(x_1,x'_1),...,$ \linebreak $P_m(x_m,x'_m) \rightarrow \bot$. \
Here, $R_i \in \mc{R}^c$, $P_j \in \mc{L}$, and $z,z' \in \bigcup \bar{x}_i \cup \bigcup \bar{y}_j$.
Some of the lists in the bodies  may be empty, i.e. $n=0$ or $m=0$, which allows to represent also classical constraints on categorical relations, e.g. keys or \fd s.


\vspace{1mm}
\noindent {\em Example:} The left-hand-side of Figure~\ref{fig:omdm} shows {\em dimensional constraint} $\eta$ on
categorical relation  {\it WorkingSchedules}, which is  linked to  the {\sf Temporal} dimension via the {\it Day} category. It says:  {\em ``No personnel was working in the Intensive care unit in January"}, i.e. \ $
\eta\!: \ \nit{WorkingSchedules}({\sf intensive},d;n,s),\nit{DayMonth}(d,$ ${\sf jan})  ~\rightarrow~ \bot$.

Dimensional \tgd \  $\sigma_1$ in Figure~\ref{fig:omdm}, given by \ $ \nit{Shifts}(w,d;n,s),\nit{WardUnit}(w,$ $u)$  $\rightarrow$  $\exists t\;\nit{WorkingSchedules}(u,d;n,t)$, says that
{\em ``If a nurse has shifts in a ward on a specific day, he/she has a working schedule in the unit of that ward on the same day"}. The use of
$\sigma_1$ generates, from the {\it Shifts} relation,  new tuples for relation {\it WorkingSchedules}, with {\em null values} for the  \nit{Specialization}  attribute, due to the existential variable. Existential rules like this (and also \egds \ and \ncs) make us depart from classic Datalog, taking us into \dpm. Relation {\em Working Schedules} may be incomplete, and new -possibly virtual- entries can be produced for it, e.g. the shaded ones showing \nit{Helen} and \nit{Sara} working for the \nit{Standard} and \nit{Intensive} units, resp.  This is done by {\em upward-navigation and data propagation} through the dimension hierarchy.
\ Constraint $\eta$ is expected to be satisfied both by the initial extensional tuples for {\it WorkingSchedules} and its tuples generated through $\sigma_1$, i.e. by its non-shaded tuples and shaded tuples in Figure~\ref{fig:omdm}, resp.
In this example, $\eta$ is satisfied.

Notice that {\it WorkingSchedules} refers to the $\nit{Day}$ attribute of the {\sf Temporal} dimensions, whereas $\eta$ involves the \nit{Month} attribute. Then, checking $\eta$ requires upward-navigation through the {\sf Temporal} dimension.  Also the {\sf Hospital} dimension  is involved in the satisfaction of $\eta$: \ The \tgd \ $\sigma_1$  may generate new tuples for {\it WorkingSchedules}, by upward-navigation from {\it Ward} to {\it Unit}.

Furthermore, we have an additional \tgd \ $\sigma_2$ that can be used with  \nit{WorkingSchedules}  to generate data for categorical relation \nit{Shifts} (the shaded tuple in it is one of them): \
$\sigma_2\!: \ \nit{WorkingSchedules}(u,d;n,t),\nit{WardUnit}(w,u)  \rightarrow \exists s\;\nit{Shifts}(w,d;n,s)$. 
It reflects the institutional guideline stating that {\em ``If a nurse works in a unit on a specific day, he/she has shifts in every ward of that unit on the same day"}. \ Accordingly, $\sigma_2$ relies on downward-navigation for tuple generation, from the {\it Unit} category level down to the {\it Ward} category level.

If we have a categorical relation ${\it Therm(Ward,Thertype};\nit{Nurse})$, with \nit{Ward} and \nit{Thertype} categorical attributes (the latter for an ${\sf Instrument}$  dimension),
the following is an \egd \ saying that {\em ``All thermometers in a unit are of the same type"}: \
${\it Therm(w,t;n)},{\it Therm(w',t';n')},\!{\it WardUnit(w,u)},\!{\it WardUnit(w',u)}$ $\rightarrow t=t'$.

 Notice that our ontological language allows us to impose a condition at the {\it Unit} level without having it as an attribute in the categorical relation.
The existential variables in dimensional rules, such as  $t$ and $s$ as in $\sigma_1$ and $\sigma_2$, resp., make up for the missing, non-categorical attributes {\it Speciality} and {\it Shift} in {\it WorkingSchedules} and {\it Shifts}, resp.\boxtheorem

 Dimensional \tgds \ can be used for  {\em upward-} or {\em downward-navigation} (or data generation) depending on the joins in the body. A one-step direction is determined by the difference of levels of the dimension categories appearing (as attributes) in the joins. Multi-step navigation, between a category and an ancestor or descendant category, can be captured through a  chain of joins with adjacent  child-parent dimensional predicates in the body of a \tgd, e.g.  propagating doctors at the unit level all the way up to the hospital level: \
$\nit{WardDoc}(\nit{ward};\nit{na},\nit{sp}),$ $\nit{WardUnit}(\nit{ward,unit}),\nit{UnitInst}(\nit{unit,ins})  \rightarrow \nit{HospDoc}(\nit{ins};\nit{na},\nit{sp})$.

\vspace{1mm}\noindent {\em Example:}  Rule $\sigma_2$ supports downward tuple-generation. When enforcing it on a tuple ${\it WorkingSchedules}(u,d;n,t)$, via category member $u$ (for Unit), a tuple for {\it Shifts} is generated {\em for each} child $w$ of $u$ in the \nit{Ward} category for which the body of $\sigma_2$ is true.
For example, chasing $\sigma_2$ with the third tuple in {\it WorkingSchedules} generates two new tuples in \nit{Shifts}: \ ${\it Shifts}({\sf W}_2, {\sf sep/6/2016},{\sf helen},\zeta)$ and ${\it Shifts}({\sf W}_1, {\sf sep/6/2016},{\sf helen},\zeta')$, with fresh nulls, $\zeta$ and $\zeta'$. The latter tuple is not shown in Figure~\ref{fig:omdm} (it is dominated by the third tuple, ${\it Shifts}({\sf W}_1, {\sf sep/6/2016},{\sf helen},$ ${\sf morning})$, in \nit{Shifts}). With the old and new tuples we obtain the answers to the query about {\it Helen}'s wards  on \nit{Sep/6/2016}: \ $\mc{Q}'(w)\!: \ \exists s\;{\it Shifts}(w, {\sf sep/6/2016}, {\sf helen},$ $ s)$.
 They are $W_1$ and $W_2$.

In contrast, the join between {\it Shifts} and {\it WardUnit} in $\sigma_1$ enables upward-navigation; and the generation of only one tuple for {\it WorkingSchedules} from each tuple in {\it Shifts}, because each {\it Ward} member has at most one {\it Unit} parent.
\boxtheorem

We can see that the \omd \ data model is an ontological model that goes far beyond classical multidimensional data models. For example, the HM model \cite{hurtado-pods}, which is subsumed by OMD, does not include general \tgds, \egds, or \ncs. Starting from our relational reconstruction of the HM model, all these elements, plus  the data and queries, are seamlessly integrated into a uniform logico-relational framework.  OMD supports general, possibly incomplete categorical relations, and not only complete ``fact tables" linked to base (or bottom) categories.

Furthermore, the constraints considered in the HM model are specific for the dimensional structure of data, most prominently, to guarantee summarizability (i.e. correct aggregation, avoiding double-counting). Specifically, we find  constraints enforcing {\em strictness} and {\em homogeneity} \cite{hurtado-pods}. The former
requires that every category elements rolls-up to a single element in a parent category, which in OMD can be expressed by \egds. The latter requires that category elements have parent elements in parent categories, which in OMD can be expressed by \tgds.   (Cf. \cite[sec. 4.3]{milaniThesis} for more details.)

 The OMD model enables {\em ontology-based data access} (OBDA) \cite{lenzerini12} and allows for the tight integration of conceptual models (e.g. an ER model expressed in logical terms) and the relational model of data, while representing and using dimensional structures and data.   Cf. \cite{milani15ruleml,journal17} for applications of the \omd \ model to quality data specification and extraction.

The ontologies of the \omd \ model have good computational properties \cite{journal17,milani15ruleml}. Actually, they belong to the class of {\em weakly-sticky} \dpm \ programs \cite{AC12}, for which conjunctive query answering (CQA) can be done in polynomial time in data. Algorithms for CQA have been proposed \cite{milani16rr,milani16rr-cali}, so as optimizations thereof \cite{milani16rr} with {\em magic-sets} techniques \cite{AL12}.

\vspace{1mm}\noindent {\bf Acknowledgements:} \ Research supported by NSERC Discovery Grant \#06148.

\end{document}